# The association of domain-specific physical activity and sedentary activity with stroke: A prospective cohort study

Xinyi He[1], Shidi Wang[2], Yi Li[1,3,4], Jiucun Wang[1,3,5,6], Guangrui Yang[7], Jun Chen[8,9,*] ,Zixin Hu[10,11,*]

[1]Human Phenome Institute, Fudan University, China

[2]Department of Social Medicine and Health Care Management, Fudan University, China

[3]State Key Laboratory of Genetic Engineering, Collaborative Innovation Center for Genetics and Development, Fudan University, China

[4]International Human Phenome Institutes, China

[5]Institute for Six-sector Economy, Fudan University, China

[6]Research Unit of Dissecting the Population Genetics and Developing New Technologies for Treatment and Prevention of Skin Phenotypes and Dermatological Diseases (2019RU058), Chinese Academy of Medical Sciences, China

[7]Exchange,Development & Service Center for Science & Technology Talents , Beijing,100045,P.R.China

[8]Sports Medicine Center, Fudan University, China

[9]Department of Sports Medicine, Huashan Hospital, Fudan University, China

[10]Artificial Intelligence Innovation and Incubation Institute, Fudan University, China

[11]Shanghai Academy of Artificial Intelligence for Science, China

* Correspondence: biochenjun@fudan.edu.cn; huzixin@fudan.edu.cn

## Abstract

### Background

The incidence of stroke places a heavy burden on both society and individuals. Activity is closely related to cardiovascular health. This study aimed to investigate the relationship between the varying domains of PA, like occupation-related Physical Activity (OPA), transportation-related Physical Activity (TPA), leisure-time Physical Activity (LTPA), and Sedentary Activity (SA) with stroke.

### Methods

Our analysis included 30,400 participants aged 20+ years from 2007 to 2018 National Health and Nutrition Examination Survey (NHANES). Stroke was identified based on the participant's self-reported diagnoses from previous medical consultations, and PA and SA were self-reported. Multivariable logistic and restricted cubic spline models were used to assess the associations.

### Results

Participants achieving PA guidelines (performing PA more than 150 min/week) were 35.7% less likely to have a stroke based on both the total PA (odds ratio [OR] 0.643, 95% confidence interval [CI] 0.523-0.790) and LTPA (OR 0.643, 95% CI 0.514-0.805), while OPA or TPA did not demonstrate lower stroke risk. Furthermore, participants with less than 7.5 h/day SA levels were 21.6% (OR 0.784, 95% CI 0.665-0.925) less likely to have a stroke. The intensities of total PA and LTPA exhibited nonlinear U-shaped associations with stroke risk. In contrast, those of OPA and TPA showed negative linear associations, while SA intensities were positively linearly correlated with stroke risk.

### Conclusions

LTPA, but not OPA or TPA, was associated with a lower risk of stroke at any amount, suggesting that significant cardiovascular health would benefit from increased PA. Additionally, the positive association between SA and stroke indicated that prolonged sitting was detrimental to cardiovascular health. Overall, increased PA within a reasonable range reduces the risk of stroke, while increased SA elevates it.

# Introduction

Stroke is a neurological condition commonly occurring in the elderly, caused by damage to brain cells due to interrupted blood supply or hemorrhage [1, 2]. It ranks as the fifth leading cause of death in the United States and is a major contributor to severe adult disability, potentially leading to serious and irreversible sequelae such as limb paralysis, speech impairment, and vision loss [3-5]. Consequently, stroke considerably affects the daily lives of patients, particularly the elderly, imposing substantial economic and living burdens on individuals, families, and society [6-8]. While stroke is typically associated with the elderly, it's gradually becoming more prevalent among younger people due to unhealthy lifestyle patterns such as inactivity, overeating, drinking, and smoking [9, 10]. Stroke treatment primarily involves medication [11, 12], interventional therapies [13], and rehabilitation training [14]. However, these strategies possess inherent limitations. For instance, while timely thrombolytic therapy [15] and interventional thrombectomy [16] are among the most effective methods, they pose certain risks to patients, including the recurrence of stroke and systemic complications. Furthermore, rehabilitation training often falls short of restoring stroke patients' physical functions to normal levels [14, 17]. Therefore, preventive measures rather than treatment should be a priority for all age groups as such measures could significantly ease the potential burdens on families and society.

Activity, a universally feasible lifestyle, can be primarily categorized into two types: Physical Activity (PA) and Sedentary Activity (SA) [18]. In recent years, the potential preventive impacts of activities on stroke have garnered significant attention from emerging epidemiological [19], clinical [20], and genetic studies [21]. However, the effects of activities on stroke derived from these studies usually contradict each other. A portion of the studies have underscored the positive influences of PA and adverse impacts of SA on stroke, whereas others have illustrated a nonlinear U-shaped relationship or a negligible connection between PA and stroke. For example, the epidemiological study of Hooker et al. [22] demonstrated that increasing moderate-vigorous PA and decreasing SA could reduce stroke prevalence. Additionally, Huan et al. [23] showed that stroke and PA shared the same genetic variants. In contrast, Rahman et al. [24] found a nonlinear U-shaped association between total PA and stroke. Moreover, Bahls et al. [21] reported that PA had limited effects on reducing stroke prevalence. Nevertheless, these studies predominantly concentrated on the total PA, neglecting the effects of various PA domains (i.e., occupation-related PA [OPA], transportation-related PA [TPA], and leisure-time PA [LTPA]) and their respective intensities (i.e., light, moderate, and vigorous) [25], which might play different roles in reducing stroke risk. As a result, studies overlooking different domains of PA and their intensities may not be able to precisely evaluate the influence of PA on stroke. As for SA, existing studies have consistently confirmed that it has a significant negative impact on stroke [21, 22, 26]. Therefore, the influences of different activities, especially different domains of PA, and their intensities on stroke remain a subject of debate.

Noting this, we collected data from the National Health and Nutrition Examination Survey (NHANES) to investigate the associations of different PA domains (i.e., OPA, TPA, LTPA, and total PA) as well as SA with stroke, respectively. First, we examined the associations between PA and SA with stroke. Then, we assessed how these associations varied across different demographic components and risk factors. Following that, we examined the dose-response relationships of PA domains and SA with stroke. Finally, we suggested feasible activity strategies to decrease stroke incidence based on the aforementioned analysis.

## Methods

### *Study design and population*

The NHANES is a continuous, national, and cross-sectional survey characterized by a sophisticated, stratified, and multistage probability sampling design, focusing on the noninstitutionalized US civilian population [27]. Considering that data regarding various levels of Physical activity (PA) and Sedentary activity (SA) were not incorporated in NHANES before 2007, and the COVID-19 pandemic outbreak in 2019 might affect the analysis results. Therefore, we utilized NHANES data collected from 2007 to 2018 as baseline data (Fig. 1). From the period of 2007 to 2018, the NHANES includes a total of 34,770 participants aged 20 years and above. The

distribution per survey cycle was as follows: 5935 in 2007-2008, 6218 in 2009-2010, 5560 in 2011-2012, 5769 in 2013-2014, 5719 in 2015-2016, and 5569 in 2017-2018.

After excluding participants who were pregnant or breastfeeding (n = 372), and those with insufficient PA and SA data (n = 168; n = 146), incomplete stroke data (n = 51), and missing covariate information (n = 3633), including variables such as education level (n = 46), annual household income (n = 2055), marital status (n = 10) and body mass index (BMI) (n = 1522), a final analytic cohort of 30,400 participants was established.

*Quantification of PA and SA*

PA and SA were evaluated based on self-reported minutes gathered by the Global Physical Activity Questionnaire (GPAQ) [28]. In this Questionnaire, participants reported the frequency, duration, and intensity of PA during a typical week. On these bases, PA is divided into three domains: occupation-related PA (OPA), transportation-related PA (TPA), and leisure-time PA (LTPA). For OPA and LTPA, their intensity is further classified as moderate and vigorous. Following the NHANES Codebook, minutes of vigorous PA were doubled and added to minutes of moderate PA to determine the total PA [29]. Subsequently, the total PA was calculated via the addition of the reported time across each domain.

To assess the influence of sufficient PA on stroke, we divided both the total PA and each domain of PA into two groups: (1) ≥ 150 min/week (meeting PA guidelines); (2) 0–149 min/week (not meeting PA guidelines). That's because the 2018 PA guidelines [30] recommend adults should engage in 150 minutes of moderate-to-vigorous intensity exercise per week, with each session lasting no less than 10 minutes. To assess the dose-response associations between different domains of PA and stroke, according to previous studies [25], we classified each domain of PA and total PA into four categories: (1) 0 min/week; (2) 1-149 min/week; (3) 150-299 min/week; (4) ≥ 300 min/week. To evaluate the dose-response relationship between SA and stroke, SA was classified as low (performing SA ≤ 7.5 h/day) or high (performing SA > 7.5 h/day), based on definitions previously provided by NHANES participants [18].

*Assessment of stroke*

Stroke was defined based on self-reported previous diagnosis during a medical consultation. Participants' stroke status was determined by their response to the question, "Have you ever been told by a doctor or other health professional that you had a stroke?" Those who answered affirmatively were considered as having experienced a stroke [31].

*Covariates*

The covariates in this study included various demographic components and risk factors closely associated with stroke. Specifically, demographic components included age (20-39, 40-59, 60-79, and ≥ 80 years), sex (male and female), race/ethnicity (Mexican American, Other Hispanic, Non-Hispanic White, Non-Hispanic Black, and Other Race), educational level (less than high school, high school, some college, and college degree or more), annual household income (0–$24,999, $25,000–$74,999, and ≥ $75,000), and marital status (never married, married/living with partner, and widowed/divorced/separated). Risk factors were BMI (low to normal, overweight, and obese) [32-34] and smoking status (nonsmoker and smoker) [32, 35].

*Statistical Analysis*

To ensure accurate representation of the broader population [31, 36], we followed guidelines from the Centers for Disease Control and Prevention (CDC), which suggested that all statistical analyses for multi-stage cluster survey design data (e.g., NHANES) should appropriately incorporate sampling weights. More specifically, the sampling weight of NHANES in this study equaled the original weight divided by the number of cycles. To summarize the participant characteristics, the categorical variables were presented as counts and corresponding weight proportions, while continuous variables as means with standard deviations. To evaluate the difference among baseline characteristics across each stroke group, we performed Rao-Scott $\chi^2$ tests for categorical variables and the Wilcoxon rank sum test for continuous variables.

In this study, PA and SA were included as exposures in the weighted multivariable logistic regression model, with stroke status as the outcome. The odds ratio (OR) and 95% confidence interval (CI) were estimated for the association between PA and SA with stroke. We conducted four distinct multivariable logistic regression models. Model 1 was the crude model. Model 2 included the Model 1 variables plus age and sex. Model 3 included the Model 2 variables plus race, education, income, marital status, and BMI. Model 4 included the Model 3 variables plus smoke status.

To further investigate the relationship between PA and SA with stroke, a subgroup analysis was utilized to explore potential variations in the impact of PA and SA on stroke across different population groups. The analysis involved various strata defined by age, sex, race, education, income, marital status, BMI, and smoking status. Furthermore, we employed meta-regression analysis to examine the heterogeneity among subgroups.

To assess the impact of various levels of PA and SA on stroke, multivariable logistic regression analysis was conducted with 0 min/week and ≤ 7.5 h/day as the reference states, respectively. To explore the potential non-linear relationships between PA and SA duration with stroke status, restricted cubic spline (RCS) was utilized in a weighted logistic model. All statistical analyses were conducted using R (version 4.2.1) software. All hypothesis tests were two-sided, and the P-value < 0.05 was considered statistically significant.

## Results

### *Population characteristic*

In our research, we gathered a dataset comprised of 30,400 participants from NHANES 2007-2018 (Fig. 1), representing 318.14 million non-institutionalized residents in the United States, with a mean (SE) age of 47.52 (16.88) years. Among 29,442 participants with missing data, 29.55% were Non-Hispanic White, 23.28% were Non-Hispanic Black, and 22.04% were Mexican American (Additional file 1: Table S1). Over the past decade, we found that the prevalence of stroke increased non-linearly across each age stratum within every period (Additional file 1: Table S2). The prevalence among females was higher than males across almost all years (Additional file 1: Table S3).

Table 1 presents the characteristics of participants categorized by stroke status. Overall, 1,166 (2.85%) participants had been informed by doctors or other health professional of their diagnosis of stroke. As compared to participants with non-stroke, those with stroke were more likely to be older, female, non-Hispanic white, married/living with partner, smoker, with higher BMI as well as lower education level and income. Moreover, according to 2018 PA guidelines [30], 18,197(64.88%) participants achieved the recommendation of total PA (performing PA ≥ 150 min/week). Furthermore, the participants who met the recommendation of OPA, TPA, and LTPA were 10,300 (37.65%), 4,206 (12.70%), and 9,962 (37.79%) respectively. For sedentary behavior, according to the previous study, 9,982 (36.16%) participants met this condition when the daily sedentary activity was more than 7.5 hours as high SA. When compared to participants with non-stroke, the prevalence of stroke participants who achieved the recommendation for total PA, OPA, TPA, and LTPA decreased in the population. Meanwhile, the proportion of participants with high SA increased among stroke participants.

*Associations of PA and SA with stroke*

We performed weighted multivariable logistic regression analyses to evaluate the association between various domains of PA and SA with stroke risk. Particularly, PA was categorized into two groups based on the PA guidelines: those who achieved the standards and those who did not. Similarly, SA was categorized as either low SA or high SA, with 7.5 h/day serving as the cutoff point (Table 2).

In Model 1, we found that total PA (odds ratio [OR] = 0.37, 95% confidence interval [CI] = [0.30, 0.44]), OPA (OR = 0.57, 95% CI 0.47-0.71), TPA (OR 0.57, 95% CI 0.40-0.80), and LTPA (OR 0.36, 95% CI 0.29-0.45) that met the PA guidelines were associated with the reduced risk of stroke in the unadjusted model. Additionally, we observed that low SA (OR 0.81, 95% CI 0.69-0.95), characterized as having SA duration below 7.5 h/day, was associated with a decreased risk of stroke. After adjusting age and sex in Model 2 and further adding race, education, income, marital status, and BMI in Model 3, the results showed that only total PA, LTPA, and low SA had significant effects. Given that smoking is a risk factor for stroke, we further incorporated smoking as a factor in Model 4. After adjusting for all demographic components and risk factors, the associations of total PA (OR

0.64, 95% CI 0.52-0.79) and LTPA (OR 0.64, 95% CI 0.51-0.81) with stroke remained significant. Meanwhile, we observed that low SA (OR 0.78, 95% CI 0.67-0.93) was associated with a decreased risk of stroke. However, OPA and TPA were not associated with stroke in the multivariable models in our analyses.

*Subgroup analysis of factors impacting the associations of PA and SA with stroke*

To evaluate the consistency of the relationship between total PA, LTPA, and SA across diverse populations, we performed subgroup analyses and interaction tests stratified by age, sex, race, education, income, marital status, BMI, and smoking (Fig. 2). For total PA, significant associations of total PA with stroke were found in each stratum. For LTPA, significant associations of LTPA with stroke were found in age, sex, race, and smoking stratum. For SA, significant associations with stroke were found in age, sex, education, income, and marital status stratum, respectively. However, the interaction tests comparing the odds ratio (OR) across the strata were not significant (P-value > 0.05), indicating that total PA and LTPA that met PA guidelines, as well as SA of less than 7.5 h/day, were associated with a lower risk of stroke, regardless of age, sex, race, education, income, marital status, and smoking status.

*Dose-response analysis of PA and SA with stroke*

To assess the potential dose-response relationships between various PA domains and stroke, as well as between SA and stroke, we categorized PA into four groups (0, 1-149, 150-299, and ≥ 300 min/week) and SA into two groups (≤ 7.5h/day and > 7.5h/day) [18]. Additionally, we evaluated the additional benefits of PA beyond or below the PA guidelines and low or high SA (Fig. 3). Overall, we found similar inverse associations across total PA, LTPA groups, and stroke. Meanwhile, direct associations across SA groups and stroke were detected. After adjusting for covariates, participants who performed total PA <1 time (1-149 min/week), 1-2 times (150-299 min/ week), or over two times (≥300 min/week) the recommended level of PA guidelines had 43% (OR 0.57, 95% CI, 0.44-0.74), 46% (OR 0.54, 95% CI, 0.38-0.76), and 45% (OR 0.55, 95% CI, 0.44-0.69) lower risks for stroke, respectively. Besides, participants with SA higher than 7.5 h/day had 28% (OR 1.28, 95% CI, 1.08-1.50) increased risk of stroke. These findings suggested that total PA, LTPA, and low SA were associated with low risks of stroke at each level. In addition, the linear trend tests for these associations were all statistically significant, with all P-values for the trend being less than 0.01.

To undercover the associations of the intensities of PA, PA domains, and SA with stroke, we further performed the multivariable-adjusted RCS analyses. For the intensities of total PA and LTPA, we observed their nonlinear U-shaped associations with stroke ($p$ for nonlinearity < 0.001; Fig. 4). Particularly, the OR initially decreased and then increased as the intensities of PA and LTPA increased, with the inflection points at 2876 and 675 minutes respectively (Fig. 4). In contrast, the intensities of OPA and TPA displayed negative linear associations with stroke ($p$ for nonlinearity = 0.462; $p$ for nonlinearity = 0.051; Fig. 4). In other words, the OR exhibited a decreasing trend with the increscent of the intensities of OPA and TPA. For the intensity of SA, we observed its positive linear association stroke, where OR increased as the intensity of SA increased ($p$ for nonlinearity = 0.904; Fig. 4).

## Discussion

The potential prophylactic impact of activities on stroke prevention has received significant scholarly attention. However, there is limited research focusing on the effects of its various domains and intensities on stroke. To scrutinize the relationship between disparate domains of Physical Activity (PA) and stroke, and the association between Sedentary Activity (SA) and stroke, we facilitated a cross-sectional analysis rooted in a large-scale cohort of 30,400 participants from the National Health and Nutrition Examination Survey (NHANES).

For PA, our findings revealed that meeting the PA guidelines for both total PA and LTPA was significantly associated with a lower prevalence of stroke. This association remained significant even after adjusting for demographic components and stroke-related risk factors. However, the link between OPA and TPA with stroke became almost insignificant following the adjustment for confounding factors. Regarding SA, the result indicated that spending over 7.5 h/day in SA significantly increased the risk of stroke. Accordingly, engaging in activities, particularly LTPA and SA, within a reasonable range, could effectively lower the risk of stroke.

In this study, the stroke prevalence among the population estimated through our data derived from NHANES was approximately 3%. Moreover, we found that stroke exhibited an unequal prevalence in different-age populations, with 70% of the participants being aged 60 years or older. This result is consistent with previous research [37-40], underscoring the importance of stroke prevention for the elderly. Additionally, our results showed a high likelihood of stroke prevalence among female, non-Hispanic white, less educated, non-living alone, smoking, and high BMI participants. These findings are consistent with previous studies, suggesting that adverse socioeconomic backgrounds [41], low education [42], smoking [35], and high BMI [43] could be associated with increased stroke risk. Intriguingly, our results revealed that compared to the non-stroke population, the stroke population is less likely to meet the recommended PA guidelines across different PA domains (i.e., OPA, TPA, and LTPA). Additionally, the stroke population tends to take SA more than 7.5 h/day compared to the non-stroke population. These results suggest the potential association between activities (i.e., PA and SA) and stroke.

From the perspective of total PA, our results reconfirmed its association with stroke [20, 22, 26], which indicates that PA could reduce the risk of stroke. From the perspective of PA domains, we found that all the PA domains (i.e., OPA, TPA, and LTPA) were associated with stroke. However, after eliminating potential confounding factors, only LTPA maintained a positive effect on reducing stroke risk. This indicates that different PA domains indeed have different influences on stroke, which is compatible with prior studies. For LTPA, numerous studies have reported its positive effect on stroke [44-48]. For example, the study of Federico et.al found that all levels of LTPA can be beneficial for stroke prevention, including levels currently regarded as low or insufficient [48]. The Asma et.al showed an inverse dose-response relationship between LTPA and the risk of stroke [47]. Therefore, our results, which are consistent with previous studies, reconfirmed the protective effects of LTPA. For OPA and TPA, our results showed their non-significant relationship with stroke. Although the findings align with previous studies, some of the others reported the opposite. For instance, Asma et.al showed that OPA was not beneficial for stroke [47]. However, Hall et.al found an inverse association between OPA and stroke [49], while Hu et.al revealed a non-inverse association [50]. We suspect that these inconsistencies could be attributed to the disparity of sex and marital status and our adoption of the new guidelines for PA intensity categorization [30]. Given that some PA domains exhibited no significant impact on stroke, such insignificant effects could overshadow the total effect of PA when all its domains are aggregated into a single variable for analysis. This might explain why certain studies did not observe the association between total PA and stroke [21, 51]. In contrast to PA, the hazardous effect of SA on stroke uncovered by our study is almost not contradicted by other studies [22].

Through subgroup analysis, we identified potential interactions between activities and confounding factors (i.e., age, sex, age, sex, race, education, income, marital status, BMI, and smoking status), as well as their collective impact on stroke. Particularly in participants of advanced age or those with high BMI, the association between activities (i.e., total PA, LTPA, and SA) and stroke became more pronounced. This suggests the heightened significance of the activities-stroke connection in these contexts. Moreover, this also demonstrates that it is necessary to adjust these confounding factors when assessing the association between activities and stroke.

The results of restricted cubic spline (RCS) quantified dynamic effects of PA, PA domains, and SA on stroke with the increscent of their intensities. First, the RCS results indicated that total PA and LTPA displayed nonlinear U-shaped associations with the increscent of their intensities. This suggests that PA would exhibit the protective effect for stroke within an appropriate range, while excessive PA may promote the risk of stroke. In contrast, SA exhibited a linear association with stroke with the increscent of its intensity. Particularly, the adverse effects on health will further intensify once the duration exceeds about 300 minutes per day. Our findings align with results from earlier studies [52-55]. Therefore, based on our research findings, we recommend engaging in a total of 255 to 2,876 minutes of PA per week to reduce the risk of stroke. Specifically, for LTPA, we suggested 0 to 675 minutes of vigorous and moderate intensity PA per week to further promote the reduction of stroke risk.

Our analysis possesses various merits. First, it delved into the associations between Physical Activity (PA) and Sedentary Activity (SA), and the incidence of stroke. Our study prudently fills this gap, since the present-day perspective of the impacts of different activity domains and their intensities on stroke remains nebulous. Our results indicated that LTPA and SA were associated with stroke prevalence across a wide age spectrum in adults, while no significant associations were found with OPA or TPA. Furthermore, total PA and LTPA exhibited non-linear trends over time, highlighting the importance of moderate-intensity PA in stroke prevention. Meanwhile, the linear trends in SA suggested that prolonged sitting could be a risk factor for stroke. Second, this study was based on a nationally representative population survey design, allowing it to more accurately represent the broader US population, thereby effectively enhancing the generalizability of its research findings. Third, novel stratified analyses eliminated the bias caused by confounding factors in the associations between exercise and stroke. Finally, we provided valuable insights for stroke prevention through trend analysis of stroke prevalence over time.

Despite these strengths, several limitations within our current study deserve further improvement. First, our study was limited by its retrospective and cross-sectional design nature, leading to information bias caused by missing data and posing challenges in establishing a causal relationship between PA and stroke, as well as SA and stroke. Second, the assessment of PA, SA, and stroke is based on self-reporting, which may have deviations from the actual situation, necessitating a more refined evaluation using clinical indicators. Our study highlights the importance of activities and suggests that interventions aiming at increasing PA and reducing SA could have a significant impact on stroke. Further Mendelian randomization analyses and mechanistic studies focusing on different domains and intensities of activities are warranted to unravel the concerted effects of various activities in the complex pathophysiology of stroke. Nevertheless, we still anticipate that our study could shed light on the promoted or prevented effects of activities and provide a reasonable activity strategy to better prevent stroke.

## Conclusions

In conclusion, our study underscores the immense potential of elevating physical activity levels and minimizing sedentary activity to confer favorable vascular effects and decelerate the progression of stroke. These results provide a plausible explanation for the complex association between lack of increasing PA and reducing SA in lowering stroke risk. Meanwhile, our recommended range of effective PA duration is 255 to 2,876 minutes per week. OPA and TPA have no restrictions, while a weekly duration of 0 to 675 minutes is advised. Furthermore, prolonged sitting might harm cardiovascular health, SA should not last more than about 300 minutes. In summary, lifestyle interventions targeting stroke can help reduce stroke risk and severity. Future research should focus on exploring the causal relationship and elucidating the precise mechanisms underlying the association between various types and intensities of activity with stroke.

## Abbreviation

PA: Physical Activity; SA: Sedentary Activity; OPA: Occupation-related physical activity; TPA: Transportation-related PA; LTPA: Leisure-time PA; NHANES: National Health and Nutrition Examination Survey; Centers for Disease Control and Prevention: CDC; BMI: Body mass index; US: United States; OR: Odds ratio; CI: Confidence interval; RCS: Restricted cubic spline.

## Supplementary Information

### Acknowledgments

We express our gratitude to the participants and team members of the National Health and Nutrition Examination Survey, as well as to the National Center for Health Statistics, for their significant contributions.

### Author contributions

XH: Writing—original draft, searched database, methodology, formal analysis, wrote this manuscript; SW: methodology, writing; YL: methodology; JW: methodology; GY: methodology; JC: supervision, writing—review & editing; ZH: supervision, writing—review & editing. All authors read and approved the final manuscript.

### Funding

Jun Chen in this study was supported by grants from the Shanghai Committee of Science and Technology (23410713100 to J.C.). Zixin Hu in this study was partially supported by funding from the National Natural Science Foundation of China (32100510 to Z.H.), and the Shanghai Rising-Star Program (21QB1400900 to Z.H.)

### Availability of data and materials

The National Health and Nutrition Examination Survey dataset is publicly available at the National Center for Health Statistics of the Center for Disease Control and Prevention (https://www.cdc.gov/nchs/nhanes/index.htm).

## Declarations

**Ethics approval and consent to participate**

NHANES is conducted by the Centers for Disease Control and Prevention (CDC) and the National Center for Health Statistics (NCHS). The NCHS Research Ethics Review Committee reviewed and approved the NHANES study protocol. All participants signed written informed consent.

**Consent for publication**

Not applicable.

**Competing interests**

The authors declare that they have no competing interests.

# Tables and Figures

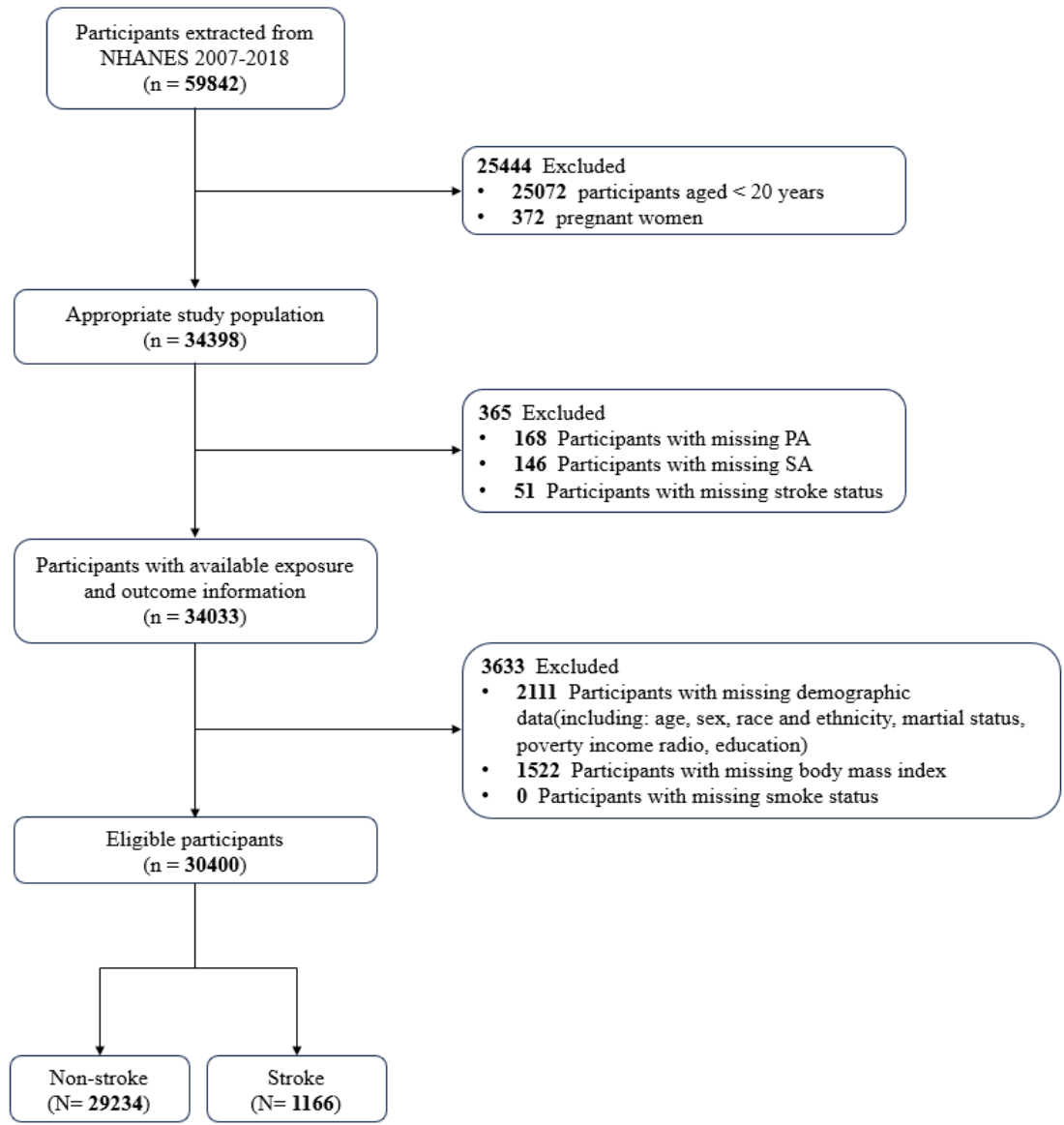

**Fig. 1** Selection of participants in the study.

PA, Physical Activity; SA, Sedentary Activity; NHANES, National Health and Nutrition Examination Survey.

**Table 1** Demographics and comparisons by stroke status.

| Characteristic | Total | Non-stroke | Stroke | P-value |
|---|---|---|---|---|
| **Sample, n (%)** | 30400 | 29234 (97.146) | 1166 (2.854) | |
| **Age, years, M (SD)** | 47.524 (16.881) | 47.028 (16.705) | 64.421 (13.859) | **<0.001*** |
| 20-39 | 10036 (36.205) | 9981 (37.096) | 55 (5.856) | **<0.001*** |
| 40-59 | 10133 (37.539) | 9870 (37.848) | 263 (27.033) | |
| 60-79 | 8354 (21.989) | 7746 (21.259) | 608 (46.845) | |
| ≥80 | 1877 (4.267) | 1637 (3.797) | 240 (20.266) | |
| **Sex, n (%)** | | | | **0.013*** |
| Male | 14859 (48.568) | 14297 (48.724) | 562 (43.237) | |
| Female | 15541 (51.432) | 14937 (51.276) | 604 (56.763) | |
| **Race/ethnicity, n (%)** | | | | **<0.001*** |
| Mexican American | 4424 (8.239) | 4325 (8.354) | 99 (4.322) | |
| Other Hispanic | 3061 (5.590) | 2990 (5.671) | 71 (2.861) | |
| Non-Hispanic White | 12697 (67.111) | 12130 (67.052) | 567 (69.105) | |
| Non-Hispanic Black | 6507 (11.086) | 6173 (10.955) | 334 (15.539) | |
| Other Race | 3711 (7.974) | 3616 (7.968) | 95 (8.174) | |
| **Education, n (%)** | | | | **<0.001*** |
| Less than high school | 7304 (15.471) | 6919 (15.181) | 385 (25.342) | |
| High school | 6953 (23.016) | 6624 (22.773) | 329 (31.277) | |
| Some college | 9015 (31.353) | 8713 (31.516) | 302 (25.802) | |
| College degree or more | 7128 (30.160) | 6978 (30.530) | 150 (17.579) | |
| **Income, n (%)** | | | | **<0.001*** |
| 0–$24,999 | 8577 (19.347) | 8037 (18.786) | 540 (38.455) | |
| $25,000–$74,999 | 12343 (39.616) | 11899 (39.576) | 444 (40.980) | |
| ≥$75,000 | 9480 (41.037) | 9298 (41.638) | 182 (20.565) | |
| **Marital status, n (%)** | | | | **<0.001*** |
| Never married | 5540 (18.312) | 5438 (7.597) | 102 (7.597) | |
| Married/Living with partner | 18034 (63.187) | 17450 (55.233) | 584 (55.233) | |
| Widowed/Divorced/Separated | 6826 (18.501) | 6346 (37.170) | 480 (37.170) | |
| **BMI, kg/m2, n (%)** | | | | **0.004**** |
| Low to normal( < 25) | 8723 (29.550) | 8424 (29.664) | 299 (25.681) | |
| Overweight(25-29.99) | 9960 (32.825) | 9594 (32.902) | 366 (30.212) | |
| Obese(≥30) | 11717 (37.625) | 11216 (37.435) | 501 (44.107) | |
| **Smoking, n (%)** | | | | **<0.001*** |
| Nonsmoker | 16920 (55.788) | 16467 (56.244) | 453 (40.266) | |
| Smoker | 13480 (44.212) | 12767 (43.756) | 713 (59.734) | |
| **Total PA: achieved, n (%)** | 18197 (64.876) | 17757 (65.576) | 440 (41.047) | **<0.001*** |
| **OPA: achieved, n (%)** | 10300 (37.650) | 10026 (37.993) | 274 (25.978) | **<0.001*** |
| **TPA: achieved, n (%)** | 4206 (12.696) | 4116 (12.843) | 90 (7.690) | **0.002**** |
| **LTPA: achieved, n (%)** | 9962 (37.792) | 9780 (38.361) | 182 (18.432) | **<0.001*** |
| **SA: high, n(%)** | 9982 (36.157) | 9519 (36.015) | 463 (40.979) | **0.010*** |

Data from National Health and Nutrition Examinations, 2007 through 2018.

PA, Physical Activity; OPA, Occupation-related physical activity; TPA, Transportation-related PA; LTPA, Leisure-time PA; SA, Sedentary Activity.

Continuous variables with normality were presented as weighted means with associated standard errors, and variables without normality were presented as weighted median with associated interquartile range. Categorical variables were expressed as numbers and corresponding weighted proportions.

*$p < 0.05$; ** $p < 0.01$; *** $p < 0.001$.

**Table 2** Multivariable OR of stroke prevalence by level of physical activity and sedentary behavior.

|  | Event, n (%) | Model 1 | | Model 2 | | Model 3 | | Model 4 | |
|---|---|---|---|---|---|---|---|---|---|
|  |  | OR (95% CI) | P-value | OR (95% CI) | P-value | OR (95% CI) | P-value | OR (95% CI) | P-value |
| **Total PA** | | | | | | | | | |
| Achieved† | 440 (41.047) | 0.366 (0.304, 0.440) | **<0.001\*\*\*** | 0.559 (0.457, 0.683) | **<0.001\*\*\*** | 0.639 (0.519, 0.786) | **<0.001\*\*\*** | 0.643 (0.523, 0.790) | **<0.001\*\*\*** |
| Did not achieve | 726 (58.953) | 1 (Ref) | 1 (Ref) | 1 (Ref) | 1 (Ref) | 1 (Ref) | 1 (Ref) | 1 (Ref) | 1 (Ref) |
| **Occupation-related PA, OPA** | | | | | | | | | |
| Achieved† | 274 (25.978) | 0.573 (0.465, 0.705) | **<0.001\*\*\*** | 0.805 (0.646, 1.003) | **0.057** | 0.813 (0.651, 1.015) | **0.072** | 0.806 (0.646, 1.006) | **0.06** |
| Did not achieve | 892 (74.022) | 1 (Ref) | 1 (Ref) | 1 (Ref) | 1 (Ref) | 1 (Ref) | 1 (Ref) | 1 (Ref) | 1 (Ref) |
| **Transportation-related PA, TPA** | | | | | | | | | |
| Achieved† | 90 (7.690) | 0.565 (0.399, 0.800) | **0.002\*\*** | 0.757 (0.533, 1.074) | **0.122** | 0.727 (0.512, 1.031) | **0.078** | 0.729 (0.513, 1.037) | **0.083** |
| Did not achieve | 1076 (92.310) | 1 (Ref) | 1 (Ref) | 1 (Ref) | 1 (Ref) | 1 (Ref) | 1 (Ref) | 1 (Ref) | 1 (Ref) |
| **Leisure-time PA, LPA** | | | | | | | | | |
| Achieved† | 182 (18.432) | 0.363 (0.292, 0.451) | **<0.001\*\*\*** | 0.503 (0.403, 0.629) | **<0.001\*\*\*** | 0.636 (0.508, 0.796) | **<0.001\*\*\*** | 0.643 (0.514, 0.805) | **<0.001\*\*\*** |
| Did not achieve | 984 (81.568) | 1 (Ref) | 1 (Ref) | 1 (Ref) | 1 (Ref) | 1 (Ref) | 1 (Ref) | 1 (Ref) | 1 (Ref) |
| **Sedentary Activity, SA** | | | | | | | | | |
| Low ($\leq 7.5$ h/day) | 703 (59.021) | 0.811 (0.694, 0.948) | **0.010\*** | 0.831 (0.707, 0.976) | **0.027\*** | 0.777 (0.659, 0.917) | **0.004\*\*** | 0.784 (0.665, 0.925) | **0.005\*\*** |
| High (> 7.5 h/day) | 463 (40.979) | 1 (Ref) | 1 (Ref) | 1 (Ref) | 1 (Ref) | 1 (Ref) | 1 (Ref) | 1 (Ref) | 1 (Ref) |

Data are presented as proportions for categorical variables, n (%), and OR (odds ratio), 95% CI (confidence interval), and P-value.

Model 1 was the crude model. Model 2 controlled for age and sex. Model 3 included the Model 2 variables plus race, education, income, marital status, and BMI.

Model 4 included the Model 3 variables plus smoke status.

†Achieved the PA recommendation (performing ≥ 150 min/week of moderate to vigorous activity).

*$p < 0.05$; ** $p < 0.01$; *** $p < 0.001$.

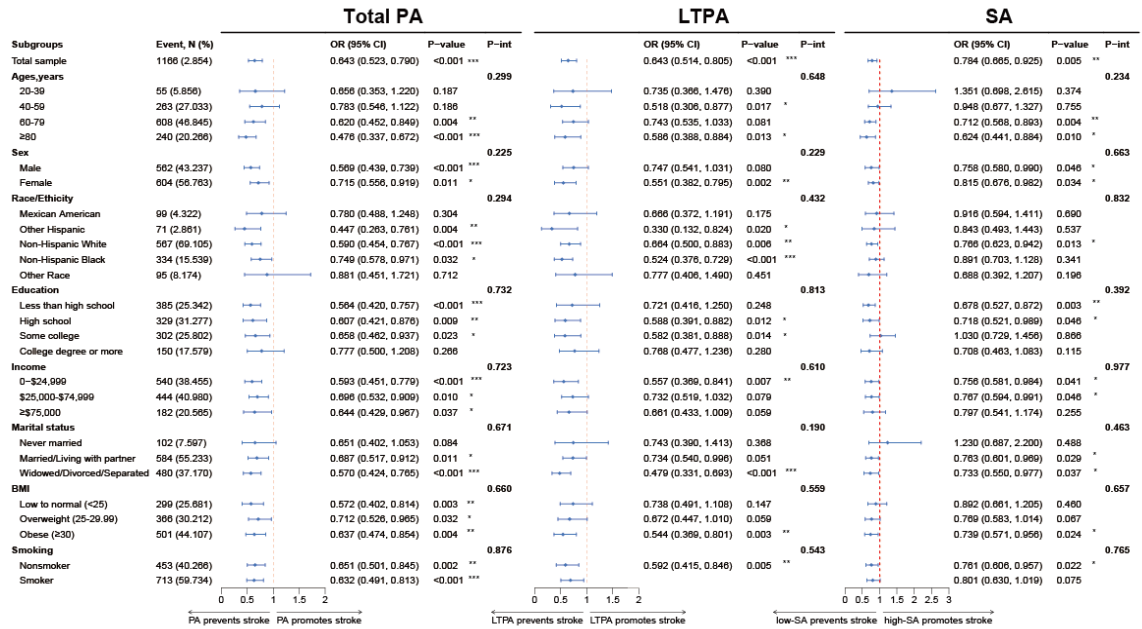

**Fig. 2** Association of total PA or LTPA achieved the PA guidelines, and SA of less than 7.5 h/day with risk of stroke in subgroups.

PA, Physical Activity; LTPA, Leisure-time physical activity; SA, Sedentary Activity; OR, odds ratio; CI, confidence interval. All ORs were adjusted for age, sex, race, education, income, marital status, BMI, and smoking status. P-int represents the heterogeneity between subgroups based on the meta-regression analysis.

*$p < 0.05$; ** $p < 0.01$; *** $p < 0.001$.

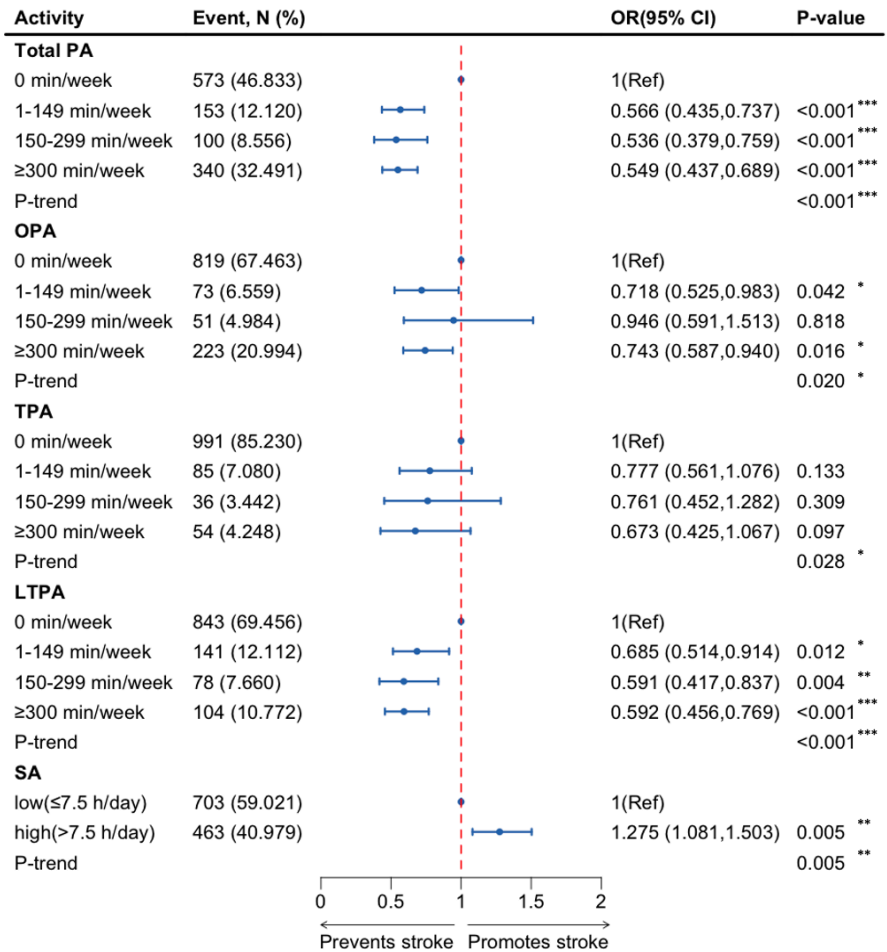

**Fig. 3** Multivariable OR for stroke based on the amount of total PA, OPA, TPA, LTPA, and SA. PA, Physical Activity; OPA, Occupation-related physical activity; TPA, Transportation-related PA; LTPA, Leisure-time PA; SA, Sedentary Activity; OR, odds ratio; CI, confidence interval. All ORs were adjusted for age, sex, race, education, income, marital status, BMI, and smoking status. *$p < 0.05$; ** $p < 0.01$; *** $p < 0.001$.

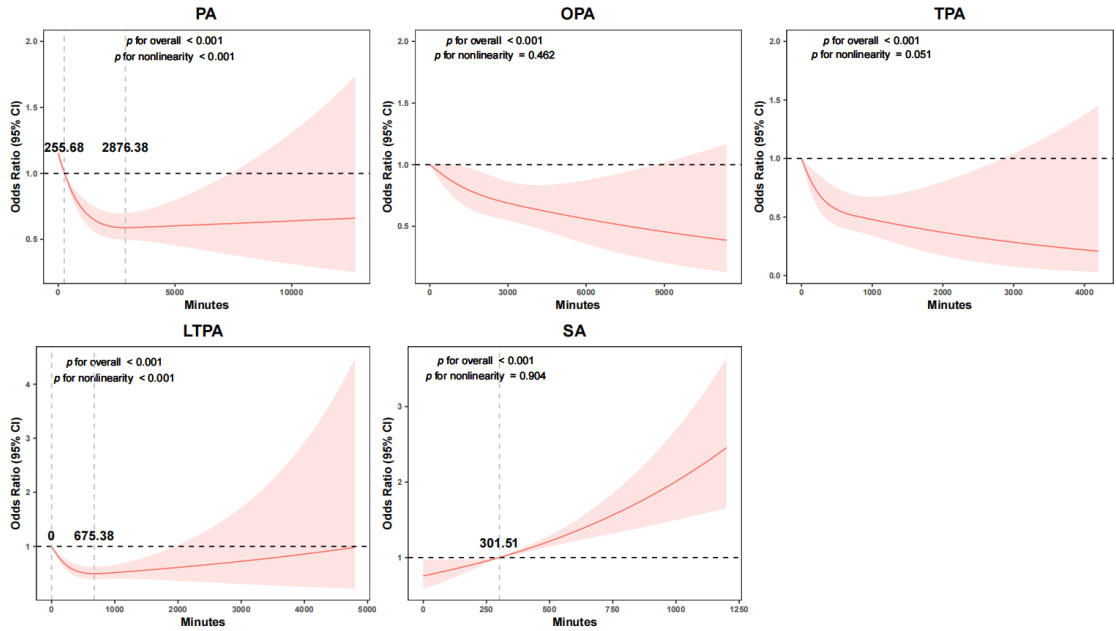

**Fig. 4** Association of PA or SA duration with stroke in a restricted cubic spline model among all participants.

Multivariable adjusted odds ratio (red solid line) with 95 % confidence interval (pink shaded area) for the association of PA minutes or SA minutes with stroke.

PA, physical activity; OPA, Occupation-related physical activity; TPA, Transportation-related PA; LTPA, Leisure-time PA; SA, Sedentary Activity; OR, odds ratio; CI, confidence interval; RCS, restricted cubic spline. All ORs were adjusted for age, sex, race, education, income, marital status, BMI, and smoking status.